# COVID-19 on YouTube: A Data-Driven Analysis of Sentiment, Toxicity, and Content Recommendations

Vanessa Su
Department of Computer Science
Emory University
Atlanta, GA 30322, USA
vanessa.su@emory.edu

Nirmalya Thakur
Department of Electrical Engineering and Computer Science
South Dakota School of Mines and Technology
Rapid City, SD 57701, USA
nirmalya.thakur@sdsmt.edu

***Abstract*—This study presents a data-driven analysis of COVID-19 discourse on YouTube, examining the sentiment, toxicity, and thematic patterns of video content published between January 2023 and October 2024. The analysis involved applying advanced natural language processing (NLP) techniques: sentiment analysis with VADER, toxicity detection with Detoxify, and topic modeling using Latent Dirichlet Allocation (LDA). The sentiment analysis revealed that 49.32% of video descriptions were positive, 36.63% were neutral, and 14.05% were negative, indicating a generally informative and supportive tone in pandemic-related content. Toxicity analysis identified only 0.91% of content as toxic, suggesting minimal exposure to toxic content. Topic modeling revealed two main themes, with 66.74% of the videos covering general health information and pandemic-related impacts and 33.26% focused on news and real-time updates, highlighting the dual informational role of YouTube. A recommendation system was also developed using TF-IDF vectorization and cosine similarity, refined by sentiment, toxicity, and topic filters to ensure relevant and context-aligned video recommendations. This system achieved 69% aggregate coverage, with monthly coverage rates consistently above 85%, demonstrating robust performance and adaptability over time. Evaluation across recommendation sizes showed coverage reaching 69% for five video recommendations and 79% for ten video recommendations per video. In summary, this work presents a framework for understanding COVID-19 discourse on YouTube and a recommendation system that supports user engagement while promoting responsible and relevant content related to COVID-19.***

## I. Introduction

The COVID-19 pandemic has been one of the most disruptive global events in recent history, significantly impacting public health, economic stability, and social structures worldwide [1]. As communities around the globe navigated waves of infection, lockdowns, and evolving public health guidelines, the flow of information about COVID-19 on the internet became crucial to shaping public understanding and response [2]. During this crisis, social media platforms emerged as essential channels for real-time information dissemination, public expression, and community interaction [3,4]. Among these platforms, YouTube has played a particularly influential role, providing a mix of authoritative updates from health organizations, personal experiences from affected individuals, and diverse user-generated commentary accessible to a global audience. The platform's video-based format uniquely facilitated comprehensive and narrative-driven insights that captured a wide spectrum of public attitudes, emotions, and interpretations surrounding COVID-19 [5,6].

As of April 2024, YouTube had more than 2.5 million monthly active users [7]. It is predicted that YouTube's audience will grow by nearly 25% over the next five years, and by 2029, YouTube's user base will comprise 1.2 billion users [8]. In 2023, YouTube accounted for 10.25% of Google's total revenue. Furthermore, in 2023, YouTube's advertising revenue was $31.5 billion, up from $29.2 billion in 2022 [9]. As of July 2024, YouTube had the most significant influence in India, with 476 million users, making it the platform's largest market. India is followed by the US (238 million users), Brazil (147 million users), Indonesia (139 million users), Mexico (84.2 million users), Japan (79.4 million users), Pakistan (66.1 million users), Germany (65.7 million users), Vietnam (63 million users), Philippines (58.1 million users), Turkey (58.1 million users), the United Kingdom (55.7 million users), and other countries [10]. The current global distribution of YouTube users shows that 54.3% are males and 45.7% are females [11]. The largest group of YouTube users worldwide are males aged 25 to 34, accounting for 12.1% of the total user base [12].

As per the findings of a survey, among Generation Z video consumers in the US, approximately 29.3% of the time they spent engaging with video format was spent on YouTube [13]. In December 2023, the highest monthly visits to YouTube, at 22 billion, were from the US. The US was followed by India (10.15 billion visits), Brazil (6.6 billion visits), South Korea (6.17 billion visits), and other countries. In December 2023 alone, YouTube recorded approximately 98 billion visits on mobile and over 8 billion visits from users on desktop devices [14]. In summary, since the beginning of COVID-19, YouTube, a globally popular social media platform, has become even more popular. Therefore, understanding the discourse on YouTube is essential for grasping the collective sentiment around the pandemic, the themes that dominate public discussion, and the nature of content that informs and influences viewers. Prior research on analyzing the public discourse about COVID-19 on social media has largely focused on text-based social media platforms like Twitter [15-17], Facebook [18-20], Instagram [21-23], and TikTok [24-26], where data is readily available and often structured around hashtag trends or keywords. Studies in this area have provided valuable insights into public sentiment, misinformation patterns, and evolving community concerns. However, video-based platforms like YouTube remain comparatively underexplored in COVID-19 research. Existing studies incorporating YouTube data are typically constrained by smaller datasets, brief time frames, or specific topics, limiting their ability to capture long-term trends and diverse thematic patterns. Moreover, most of these works lack a comprehensive analysis of sentiment and toxic content in video descriptions, dimensions essential to understanding the public's emotional response and potential exposure to negativity on YouTube.

This study addresses these gaps related to the understanding of COVID-19-related discourse by presenting a dataset of 9235 videos published publicly on YouTube between January 2023 and October 2024. After the development of this dataset, a comprehensive analysis of the video descriptions was also performed, which included natural language processing and machine learning approaches to derive sentiment scores, classify toxicity levels, and categorize thematic elements to understand how COVID-19 is discussed on YouTube and highlight the complex interplay of positive, neutral, and negative sentiments, as well as toxic elements that could impact public well-being. Beyond conventional sentiment classification, the inclusion of toxicity detection in this study is particularly relevant due to the role of toxic content on social media in shaping public attitudes during health crises [27,28]. A recommendation system was also developed specifically for COVID-19 video content on YouTube. Traditional recommendation systems often rely on user engagement metrics, such as views and likes [29,30], which may not correlate with content quality or reliability [31,32]. In contrast, the recommendation system developed in this paper prioritizes sentiment, toxicity, and thematic alignment, guiding users toward content that is relevant and less likely to be misleading or toxic. By focusing on emotional and thematic relevance, the proposed recommendation system aims to improve content discoverability and guide users through the extensive COVID-19 information landscape on YouTube. It also seeks to minimize the unintentional promotion of toxicity and contribute towards a more informed and balanced viewing experience on YouTube. In summary, this research makes three major contributions related to the study of COVID-19 discourse on YouTube. First, it introduces a large-scale dataset capturing COVID-19-related content, providing a valuable resource for analyzing pandemic-related discussions over time. Second, it conducts in-depth sentiment analysis, toxicity analysis, and topic modeling to reveal the emotional and psychological dimensions of this content, offering insights into the themes and potential impact of video-based discourse on YouTube. Third, it develops a video recommendation system that prioritizes sentiment, toxicity, and thematic alignment, guiding users toward content that is both relevant and safe, thereby enhancing the quality and reliability of video recommendations.

The rest of this paper is structured as follows. Section 2 presents a review of recent works in this field. The methodology is discussed in Section 3, followed by the results in Section 4. Section 5 concludes the paper and summarizes the findings of this research work.

## II. Literature Review

Since the beginning of the pandemic, several works have focused on the role of YouTube as a platform for disseminating information and capturing public views, opinions, perspectives, and sentiment toward COVID-19.

Obadimu et al. [33] studied how toxic behavior spread in YouTube discussions during the early days of the COVID-19 pandemic. They collected data from 544 channels that comprised more than 3,488 videos and 849,689 comments. Through topic modeling and social network analysis, they identified influential commenters and the top toxic users in the network. Jin et al. [34] examined the polarization and effects of user engagement on COVID-19-related videos on YouTube. They analyzed 6,668 videos, and the findings of their work showed that user engagement with these videos progressively increased over time. Their work also showed that videos containing lower content bias generally achieved higher rankings in search results on YouTube. Di Marco et al. [35] studied the formation of echo chambers on YouTube during the pandemic. They studied the data of 2 million users' engagement released by 68 YouTube channels. The findings showed that the echo chamber structure cannot be reproduced after properly randomizing user interaction patterns. Their study also indicated a relation between the political bias of users and their tendency to consume highly questionable news. Jagtap et al. [36] developed an approach for detecting misinformation related to COVID-19 on YouTube based on the analysis of video subtitles. They analyzed a dataset of over 2943 videos related to COVID-19. Their approach classified the videos into three classes (misinformation, debunking misinformation, and neutral) with a 0.85 to 0.90 F1 score. Triantafyllopoulos et al. [37] presented the COVYT dataset, comprising speech samples from 65 individuals extracted from platforms such as YouTube and TikTok. Some of the individuals had COVID-19, whereas others did not. Their research focused on studying the impact of COVID-19 on speech, and their work showed that it was possible to detect COVID-19 based on speech samples collected from YouTube.

Xie et al. [38] focused on the challenges faced by people with visual impairments (PVI) during the pandemic, specifically on YouTube. They studied 24 videos published by PVI and 27 videos published by the news media community, where 57 PVI were depicted. The findings of their study showed that PVI accessibility can be easily missed on YouTube. Breazu et al. [39] investigated the increasing racism and xenophobia with a particular focus on the Roma community during the COVID-19 pandemic. The authors performed a critical discourse analysis of comments on a YouTube video published early in the pandemic. The findings showed that users expressed racist and xenophobic views, particularly within the Romanian context. They also discussed how the affordances of social media might increase the chances of being exposed to racism. Soares et al. [40] studied the promotion of Hydroxychloroquine (HCQ)-related misinformation on YouTube in Brazil by focusing on two research questions: first, how pro-HCQ content is disseminated on YouTube, and second, how YouTube's algorithm recommends videos about HCQ. They used a mixed methods approach and concepts of social network analysis to investigate the content of 751 videos. Their study showed that mainstream media were the primary sources for promoting pro-HCQ videos on YouTube. Furthermore, they demonstrated that YouTube's algorithm was more likely to recommend pro-HCQ videos than anti-HCQ videos. Li et al. [41] investigated the reliability and effectiveness of the top-viewed YouTube videos about COVID-19. The first step of their study involved data mining, which was performed by collecting videos from YouTube based on keyword searches. The keywords used were "coronavirus" and "COVID-19", and out of 150 videos (top 75 viewed videos per keyword), they included 69 videos for their study. The exclusion criteria of their study included duplicate videos, non-English, non-audio, non-visual, exceeding 1 hour in duration, live, and unrelated to COVID-19. The results of their work showed that 27.5% of the videos contained misleading information about COVID-19, and those videos had a cumulative of 62042609 views. Based on the same, they also recommended better utilization of YouTube by public health agencies to disseminate accurate information about COVID-19. The work done by Suter et al. [42] focused on understanding the connection between political polarization and misinformation related to COVID-19. They analyzed 3.5 million YouTube comments on COVID-19-related videos published on four YouTube channels over 16 months. Thereafter, they classified these comments as fake or legitimate. The results showed that the proportion of fake news increased in the comments sections with time at a rate of 0.4% per month. The work also indicated that fake comments have more likes than legitimate comments. The focus of the work done by Li et al. [43] was to investigate user engagement and emotional response towards COVID-19-related videos on YouTube. They studied 38,085 YouTube comments published on the most viewed videos related to COVID-19. They used the NRC lexicon to detect the sentiments in the video titles and comments. The findings of their study showed that longer video titles and sad emotions tend to receive more likes on YouTube.

Rodriguez et al. [44] studied exercise-related videos disseminated on YouTube during COVID-19 to evaluate the compliance of the exercises in those videos with WHO recommendations. Their work involved the investigation of 150 videos, out of which 82 were eliminated before data analysis for not meeting their specified inclusion criteria. They used the principal component analysis (PCA) to classify these videos as relevant and non-relevant. They also computed the video power index of all the videos to evaluate the overall popularity of such videos on YouTube. They utilized DISCERN, Health on the Net code (HONCode), and global quality score (GQS) to assess the quality of these videos. The results of applying these approaches were 2.29, 58.95, and 2.32, respectively. Hall et al. [45] analyzed 100 comments about COVID-19 on a YouTube video about functional disability. They identified four themes - lack of access to care and services, isolation and lifestyle changes, mental health consequences, and peer support in these comments. The work of Zhang et al. [46] involved a comprehensive investigation of public views related to COVID-19, as expressed in the comments section of YouTube videos of Canadian Prime Minister Justin Trudeau's COVID-19 daily briefings. They studied a total of 46,732 English comments published on 57 videos. They identified 11 themes in these comments, which included strict border measures, public responses to Prime Minister Trudeau's policies, essential work and frontline workers, individuals' financial challenges, rental and mortgage subsidies, quarantine, government financial aid for enterprises and individuals, personal protective equipment, Canada and China's relationship, vaccines, and reopening.

Carvache-Franco [47] studied comments published on videos about tourism during COVID-19. They studied a total of 39,225 comments published in different languages. The findings of their work showed that people, country, tourist, place, tourism, see, visit, travel, covid-19, life, and live, were the most frequently discussed topics in these comments. Their study also showed that India, Nepal, China, Kerala, France, Thailand, and Europe were among the frequently mentioned destinations. Ng et al. [48] focused on investigating the health communication in YouTube videos about COVID-19 published by content creators on YouTube from the US, Brazil, Russia, Taiwan, Canada, and New Zealand. They applied the extended parallel process model to 2152 trending COVID-19-related videos in their study. The results showed that COVID-19-related videos from Taiwan promptly gained user attention on YouTube. However, such videos from the US and Brazil took considerable time to gain user attention. Keijzer et al. [49] studied the role played by YouTube in creating and disseminating COVID-19-related conspiracy theories and extreme beliefs. They applied a mixed methods approach to analyze 10,000 videos with up to 20 recommendation links per video. They manually labeled more than 33% of the videos for the analysis, which also included a clustering process to determine groups of similar videos. The findings of their study showed that YouTube's algorithm did not promote conspiracy theories and extreme beliefs. Parabhoi et al. [50] studied the role of YouTube in actively disseminating COVID-19-related information. They collected a total of 1084 videos for their study. Thereafter, they excluded videos unrelated to COVID-19 or published in a language other than English or Hindi and performed a comprehensive data analysis on the remaining 349 videos. Their study showed that most videos contained information about COVID-19; 11.17% focused on treatments, and 4.01% focused on symptoms. Their work also showed that news channels, government sources, and healthcare professionals accounted for 71.63%, 6.87%, and 5.74% of these videos, respectively. D'Souza et al. [51] analyzed the accuracy of medical information related to COVID-19 in YouTube videos. Their work involved coding 113 videos related to COVID-19 by considering each video's characteristics, source, and medical content. The findings of their work showed that the percentage of useful and misleading videos was 69.9% and 8.8%, respectively. Their study further showed that independent content creators on YouTube were more likely to post misleading content, and news agencies published more useful content than misleading content.

In summary, despite several works in this field, the literature on COVID-19 discourse on YouTube reveals limitations in scope, dataset size, and analytical depth, particularly in areas like long-term sentiment trends, toxicity detection, and thematic diversity. The work of this paper aims to address these research gaps. The step-by-step process that was followed for this research project is presented in Section III.

## III. Methodology

Figure 1 presents the step-by-step methodology that was followed in this research project. The dataset was developed by writing a program in Python 3.10 that performed data mining by connecting with the YouTube API. This program used keyword search for the data mining process, and the keywords that were used to collect COVID-19-related videos were "COVID," "COVID19" "coronavirus," "COVID-19", "corona," and "SARS-CoV-2". For each video that contained at least one of these keywords in the title or the description, the program stored the URL of the video, video ID, video title, video description, and date of publication of the video in a CSV file. The information of 15,145 videos published on YouTube between January 1, 2023, and October 25, 2024 (the most recent date at the time of performing data mining) was obtained from this program. Thereafter, duplicate records were removed, and videos that were not related to COVID-19 were deleted, which resulted in the CSV file containing the data of 9,325 videos published between the same date range. Then, the yt_dlp package in Python [52] was used to extract the number of views, likes, comments, duration (in seconds), categories, tags, language, and availability of captions per video. These results were stored as separate attributes in the dataset.

The next step, data preprocessing, was crucial for standardizing and preparing the text data for the model-building and data analysis tasks. The data preprocessing was performed on the video descriptions. The pipeline included converting text to lowercase, removing URLs, user mentions, hashtags, and non-alphabetic characters, retaining emotion-specific emojis, eliminating common English stop words, stemming, and lemmatization. Here, emotion-specific emojis refer to 😀, 😃, 😄, 😁, 😆, 🥵, 🥶, 😱, 😨, 😰, 😥, 😓, 🤔, 🤗, 🤭, 🤫, 😐, 😑, 😶, 🙄, 🤨, 🧐, 🤓, 😎, 😏, 😒, 😔, 😟, 😕, 🙁, ☹️, 😣, 😖, 😫, 😩, 😢, 😭, 😤, 😠, 😡, 👿, 💀, ☠️, 👻, 👽, 🤖, 🎃, 😺, 😸, 😹, 😻, 😼, 😽, 🙀, 😿, and 😾, as per the findings of prior work [53] in this field. Sentiment analysis of the video descriptions was conducted using the Valence Aware Dictionary and sEntiment Reasoner (VADER) [54], as it has been used in several prior works in this field [55-58]. VADER is a lexicon-based model that combines lexical scoring with heuristic rules to interpret sentiment in a given text. VADER works by parsing each word in a text and assigning it a valence score based on a predefined sentiment dictionary. Additionally, VADER accounts for common linguistic patterns that modify sentiment, such as intensifiers (e.g., "very"), negations, capitalization, punctuation, and emoticons. The output was a compound sentiment score - a composite measure capturing the overall sentiment of a text. This study used the compound sentiment score to classify videos into positive, negative, or neutral categories, with threshold values set to distinguish between the three classes. The toxicity analysis utilized Detoxify [59], a deep-learning model based on a transformer architecture. Detoxify is fine-tuned on a dataset annotated for toxic language, which enables it to detect multiple forms of toxicity, including hate speech, insults, and threats. Built on the Bidirectional Encoder Representations from Transformers (BERT) model [60], Detoxify processes text by capturing the context and interdependence of words through self-attention layers, allowing it to learn relationships across entire text sequences.

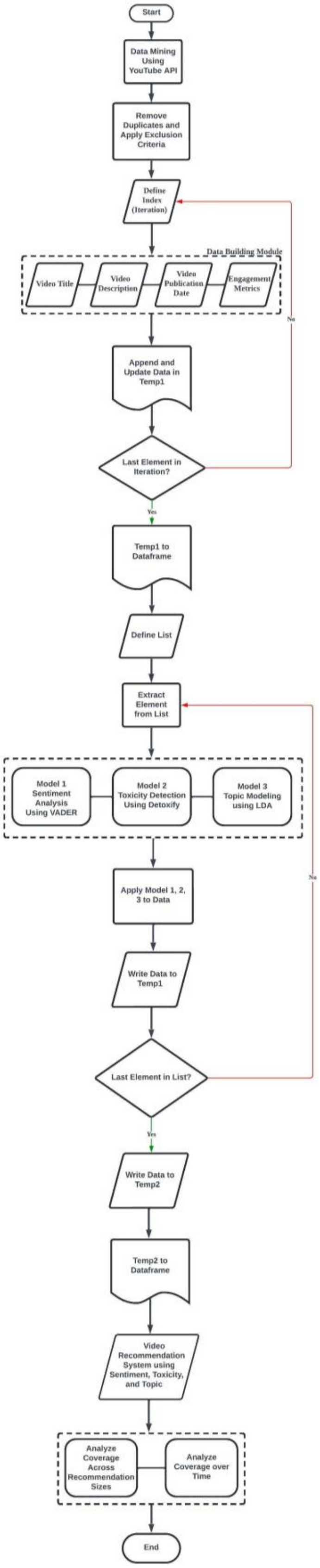

Figure 1. A flowchart that represents the methodology followed in this research project

The model outputs a toxicity probability score for each text input, estimating the likelihood that the content contains toxic language. This study applied a threshold to classify descriptions as toxic or non-toxic, measuring potentially toxic content. Thereafter, topic modeling was implemented to identify thematic patterns within the dataset using Latent Dirichlet Allocation (LDA) [61,62], a probabilistic model that assumes each document is a mixture of a small number of topics and each topic is a mixture of words. Preprocessed video descriptions were vectorized using the CountVectorizer, which converted the text into a matrix format where each row represented a document and each column corresponded to word frequency. The LDA model was applied to this matrix to infer topic distributions across descriptions. In LDA, each document's content is assumed to be generated by a distribution over latent topics, where a specific word distribution characterizes each topic. During model training, a grid search with cross-validation was used to determine the optimal number of topics, which was found to be two, balancing interpretability and thematic coherence. Once trained, the model assigned topics to videos based on their word distributions, providing insights into the primary themes present in the discourse.

Finally, to develop the recommendation system based on sentiment, toxicity, and topic modeling, a combined feature representation of each video's preprocessed title and description was created, allowing both elements to contribute to the similarity-based recommendation algorithm. Textual similarity between videos was assessed using Term Frequency-Inverse Document Frequency (TF-IDF) vectorization [63]. It transforms text data into numerical vectors by weighing terms according to their importance within a document relative to the entire corpus. By giving higher weight to distinctive terms, TF-IDF helped highlight keywords that defined each video's unique content, which was essential for accurate similarity measurements. Cosine similarity was then calculated between all TF-IDF vectors to quantify content alignment across video descriptions, creating a matrix of similarity scores for each pair of videos. For each video, the top n most similar videos were selected as recommendations and filtered further based on sentiment, toxicity, and topic classification to ensure relevance. This filtering step restricted recommendations to videos sharing the same sentiment, toxicity classification, and thematic category as the query video, aligning recommendations with viewer expectations and reducing the likelihood of exposure to toxic or negative content. Several metrics were utilized to evaluate the recommendation system's performance and robustness. Aggregate coverage, defined as the proportion of unique videos recommended across the entire dataset, was calculated to measure the system's overall diversity in content recommendations [64,65]. Coverage over time [66,67] was also computed by calculating the monthly recommendation coverage, capturing temporal trends in recommendation reliability as the dataset evolved over the 21-month period. Cumulative coverage was used to monitor the system's scalability, measuring the expansion of unique recommendations as more videos were processed. Additionally, coverage by recommendation size was assessed by evaluating the number of recommendations provided per video (top 1, top 5, top 10) and its impact on overall coverage. The combination of these metrics provided a detailed view of the recommendation system's adaptability and effectiveness across varying levels of recommendation breadth, indicating its capacity to provide relevant content while supporting thematic and sentiment consistency. To develop and implement all the models described in this section, programs were written in Python 3.10. The design of the recommendation system, particularly its emphasis on sentiment, toxicity, and thematic alignment, highlights this study's commitment to enhancing the viewer experience on YouTube by prioritizing relevance and ensuring that recommended content aligns with the viewer's informational and emotional needs. The results are presented and discussed in Section IV.

## IV. Results and Discussions

The results of this study encompass the findings of several stages of this research project, including dataset development, sentiment analysis, toxicity analysis, topic modeling, and recommendation system evaluation. The dataset that was developed is available on IEEE Dataport at https://dx.doi.org/10.21227/sbj6-pt91. This dataset contains the data of 9,325 YouTube videos related to COVID-19, with the following key attributes:

- Video URL: The complete URL of a video, for example: https://www.youtube.com/watch?v=3aHLwUD2iPg
- Video ID: The unique ID of a video, for example, 3aHLwUD2iPg.
- Title: The title of the video.
- Description: Description of the video.
- Publish Date: The date of publication of a video. The dates range from January 1, 2023, to October 25, 2024.
- View Count: The number of views of a video. It ranges from 0 to 30,107,100, with a mean of approximately 59,803 views.
- Like Count: The number of likes of a video. It ranges from 0 to 607,138 and has an average of around 1,413 likes.
- Comment Count: The number of comments on a video. It ranges from 1 to 25,000 comments, with an average of 147 comments.
- Duration (seconds): The duration of a video. Varies from 0 to 42,900 seconds, with a median duration of 137 seconds.
- Categories: There are 15 unique categories, with "News & Politics" as the most common (4,035 videos).
- Tags: Tags associated with the videos
- Language: The specific language of the video with "en" (English) as the most common.

Next, the compliance of this dataset with the FAIR (Findability, Accessibility, Interoperability, and Reusability) principles for scientific data management [68] is outlined. The dataset is findable, as it has a unique and permanent DOI, which was assigned by IEEE Dataport. The dataset is accessible online. It is interoperable due to the use of a CSV for data representation that can be downloaded, read, and analyzed across different computer systems and applications. The dataset is reusable as the associated video-related information, such as video ID, video title, video description, number of views, likes, comments, duration (in seconds), categories, tags, and language per video, can be reused any number of times for any application. Here, the compliance of this dataset with the FAIR principles is discussed as several prior works related to data mining that led to the development of datasets such as RCSB Databank [69], Pdebench [70], Wiki pathways [71], the open reaction dataset [72], MGnify [73], MiMeDB [74], HMDB 5.0 [75], the Pfam database [76], and datasets of Tweets about COVID-19 [77-80], outlined how the associated datasets complied with the FAIR principles of scientific data management.

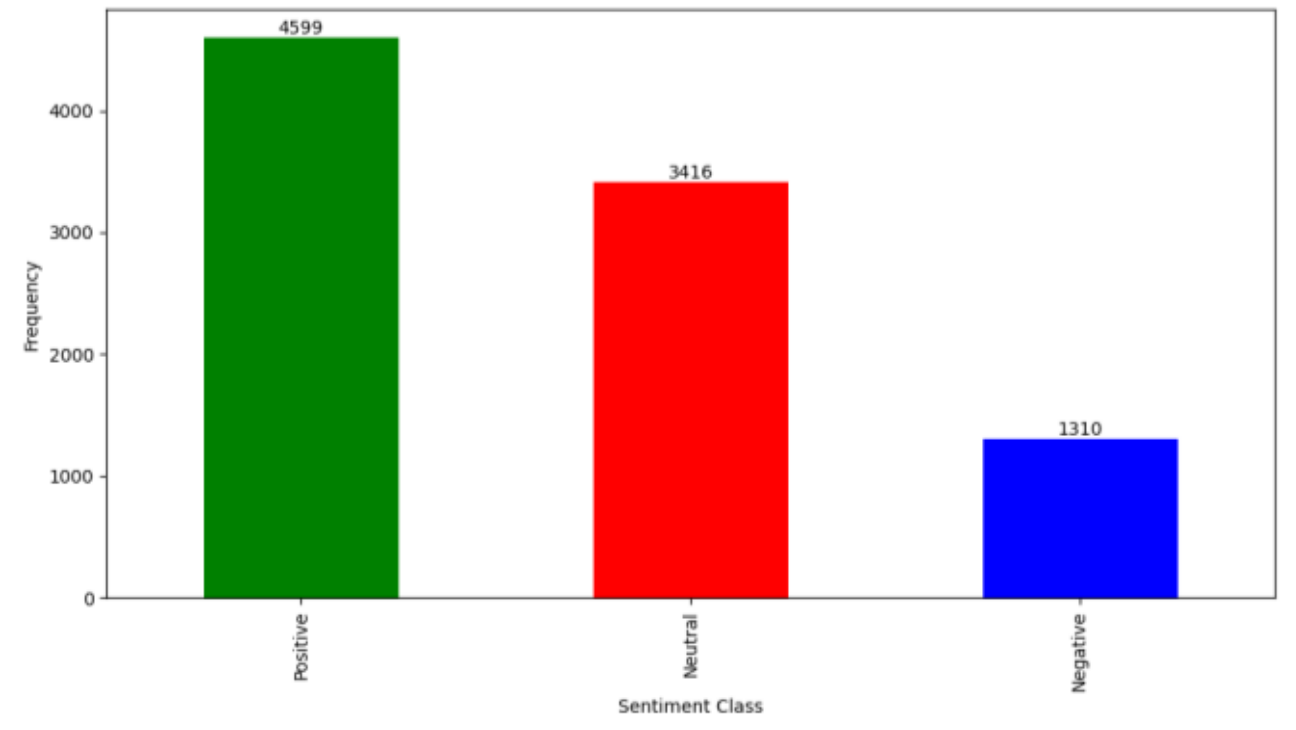

Figure 2. Results of performing Sentiment Analysis of the video descriptions

The results of sentiment analysis of the video descriptions, conducted using VADER, are shown in Figure 2. It was observed that 4,599 videos, or 49.32% of the total, contained a positive sentiment. Neutral sentiments followed closely, with 3,416 videos (36.63%), while negative sentiments were least frequent, at 1,310 videos (14.05%). This trend indicates that a substantial portion of COVID-19 content on YouTube was probably aimed to inform or uplift viewers, likely reflecting creators' intentions to foster awareness and reassurance during the pandemic. The results also indicate that while negative content was present, it remained relatively limited, potentially focusing on pandemic-related challenges or news updates on risks and losses. In the toxicity analysis performed using Detoxify, it was observed that an overwhelming majority of the videos were non-toxic. Specifically, 9,240 videos (99.09%) were classified as non-toxic, while only a small subset, 85 videos (0.91%), was labeled as toxic. These results are shown in Figure 3. The results from Figure 3 indicate that most content creators maintained a non-offensive, respectful tone when discussing COVID-19 on YouTube. The low incidence of toxic content suggests that YouTube served as a relatively safe space for pandemic-related discourse, with minimal exposure to overtly negative content. However, even a small fraction of toxic content underscores the need for content moderation efforts to prevent the spread of potentially toxic content that could impact public sentiment negatively. Topic modeling using Latent Dirichlet Allocation (LDA) further segmented the video content into distinct themes. Through GridSearchCV, the optimal number of topics was identified as two (Figure 4). Topic 0, comprising 6,223 videos (66.74%), was characterized by leading terms such as "covid", "coronavirus", "health", and "pandemic," pointing to a general focus on COVID-19 discussions, including health implications and updates on the pandemic's broader impacts. Topic 1, which included 3,102 videos (33.26%), was defined by leading terms such as "news", "channel", "live", and "update", indicating a focus on real-time news coverage and updates.

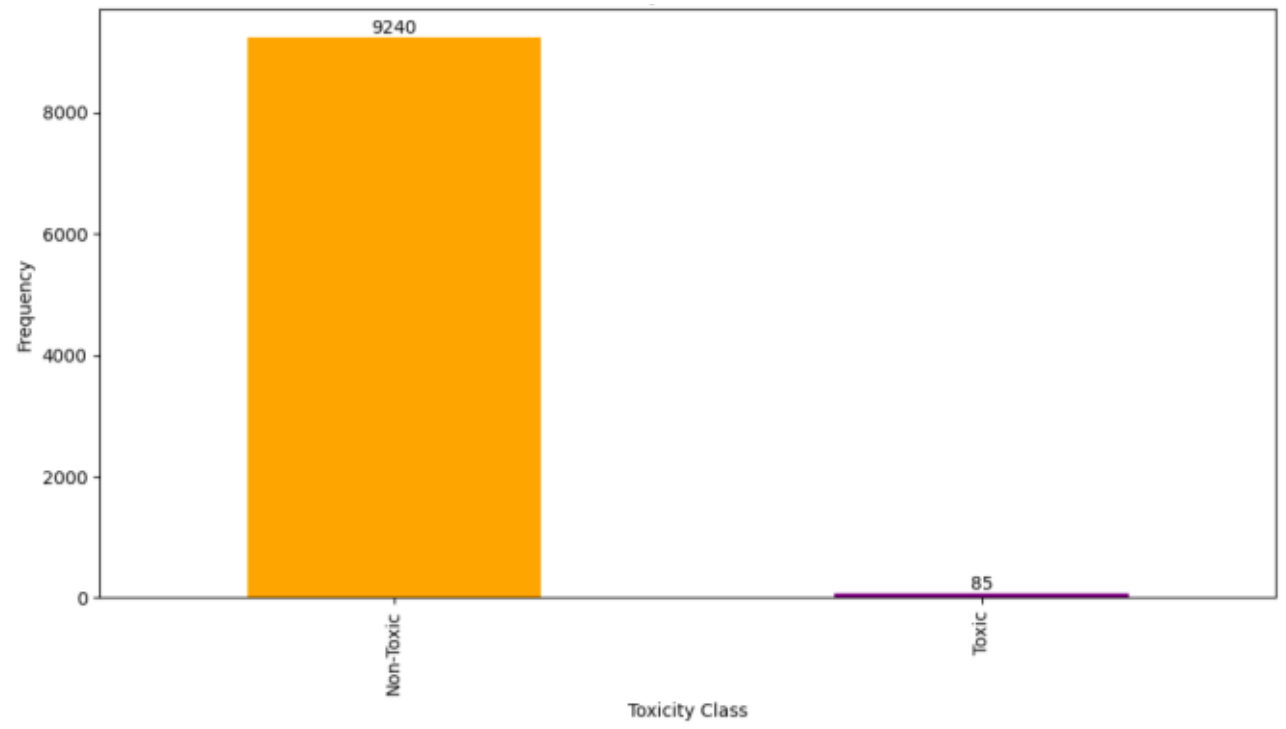

Figure 3. Results of performing Toxicity Analysis of the video descriptions

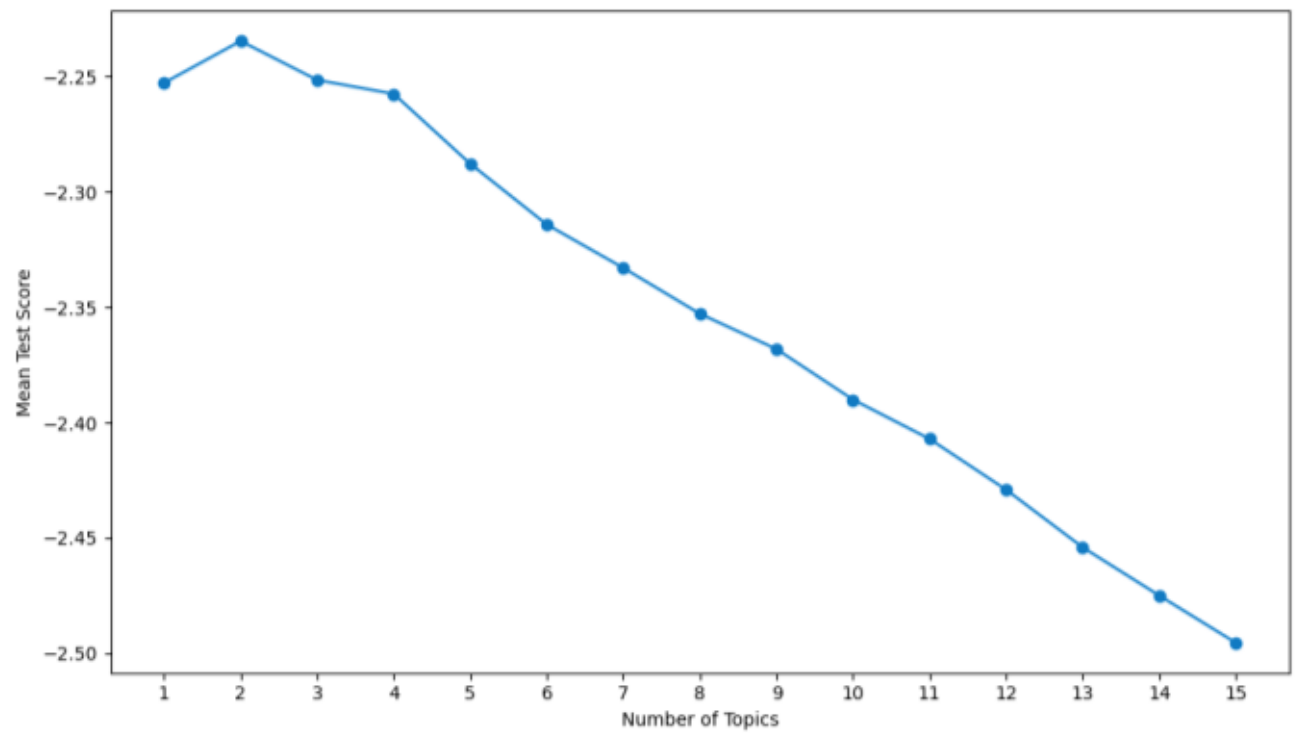

Figure 4. Determination of the optimal number of topics using GridSearchCV

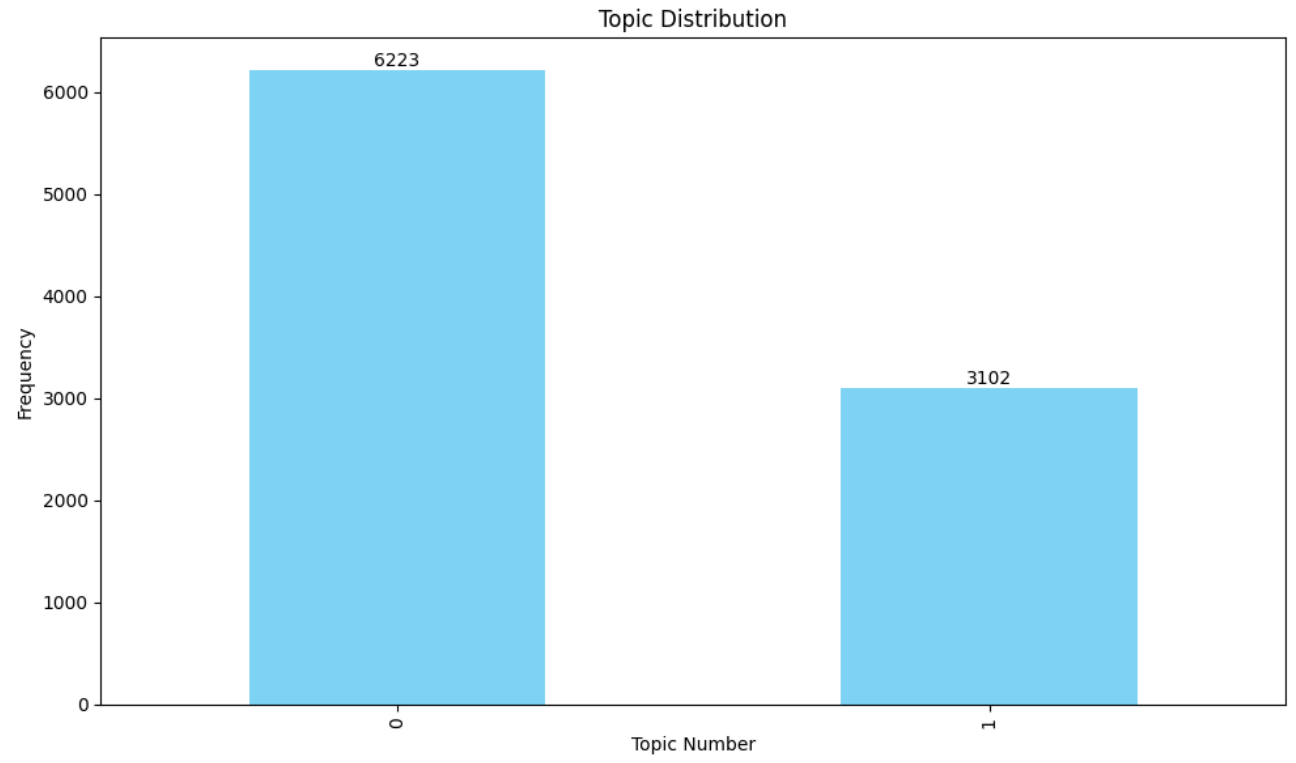

Figure 5. Distribution of videos between the two topics

Figure 5 shows the number of videos per topic. An analysis was also performed to identify the most frequent words per topic. The results are shown in Figures 6 and 7, respectively. The recommendation system was evaluated for its effectiveness in providing relevant content suggestions by using TF-IDF vectorization on video titles and descriptions, followed by cosine similarity computation. The recommendation system achieved an aggregate coverage of 0.69, meaning that 69% of videos in the dataset received at least one unique recommendation. Temporal analysis of coverage over time demonstrated consistency, with monthly coverage rates varying between 0.83 in October 2024 and peaking at 0.93 in December 2023. These findings suggested that the recommendation system maintained relevance and adaptability across different pandemic phases. To further assess the recommendation system's robustness, cumulative coverage was calculated by tracking the accumulation of unique recommendations as videos were incrementally processed. The cumulative coverage metric reached a maximum of 0.69, demonstrating that the recommendation system scaled effectively with the dataset size, indicating its potential applicability to larger video collections. Additionally, an analysis based on recommendation size revealed that single recommendations (top 1) achieved coverage of 0.37, whereas expanding to five recommendations (top 5) increased coverage to 0.69 and offering ten recommendations (top 10) further raised coverage to 0.80. These results are shown in Figures 8 and 9, respectively.

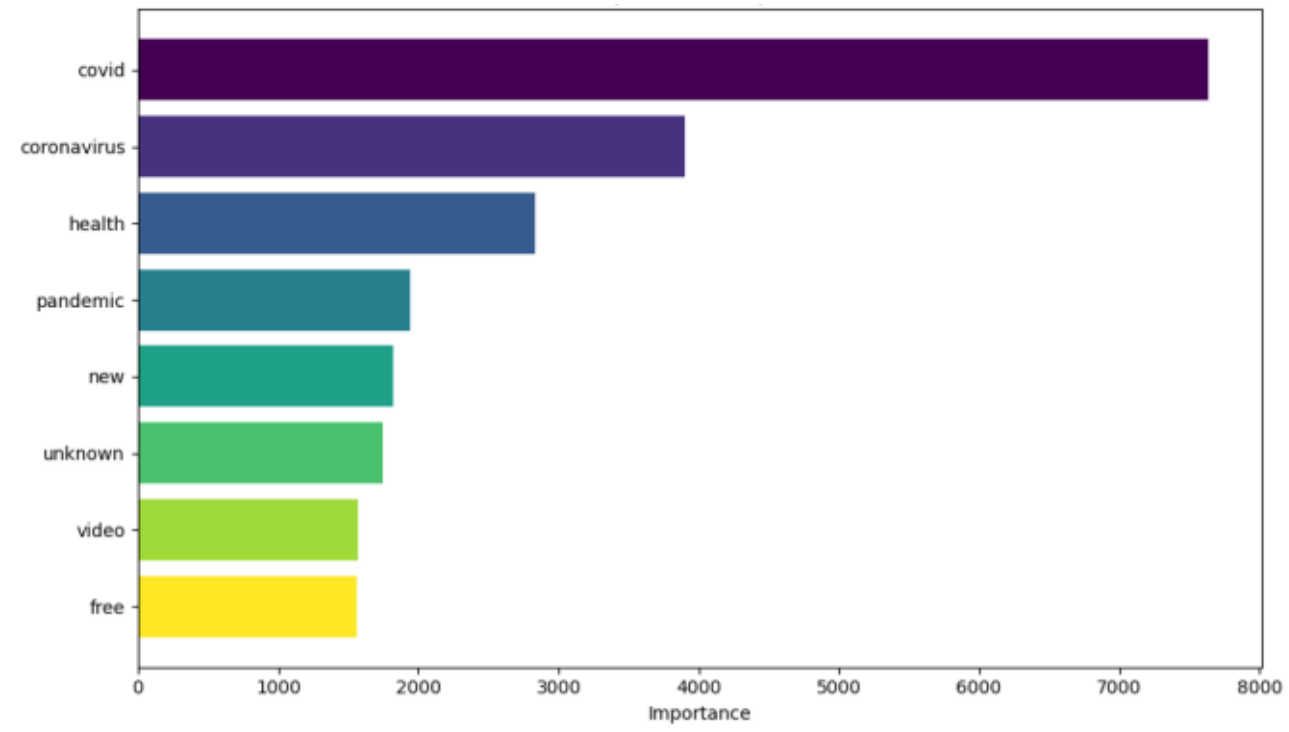

Figure 6. The top 8 frequent words in Topic 0 that focused on COVID-19 discussions, including health implications and updates on the pandemic's broader impacts

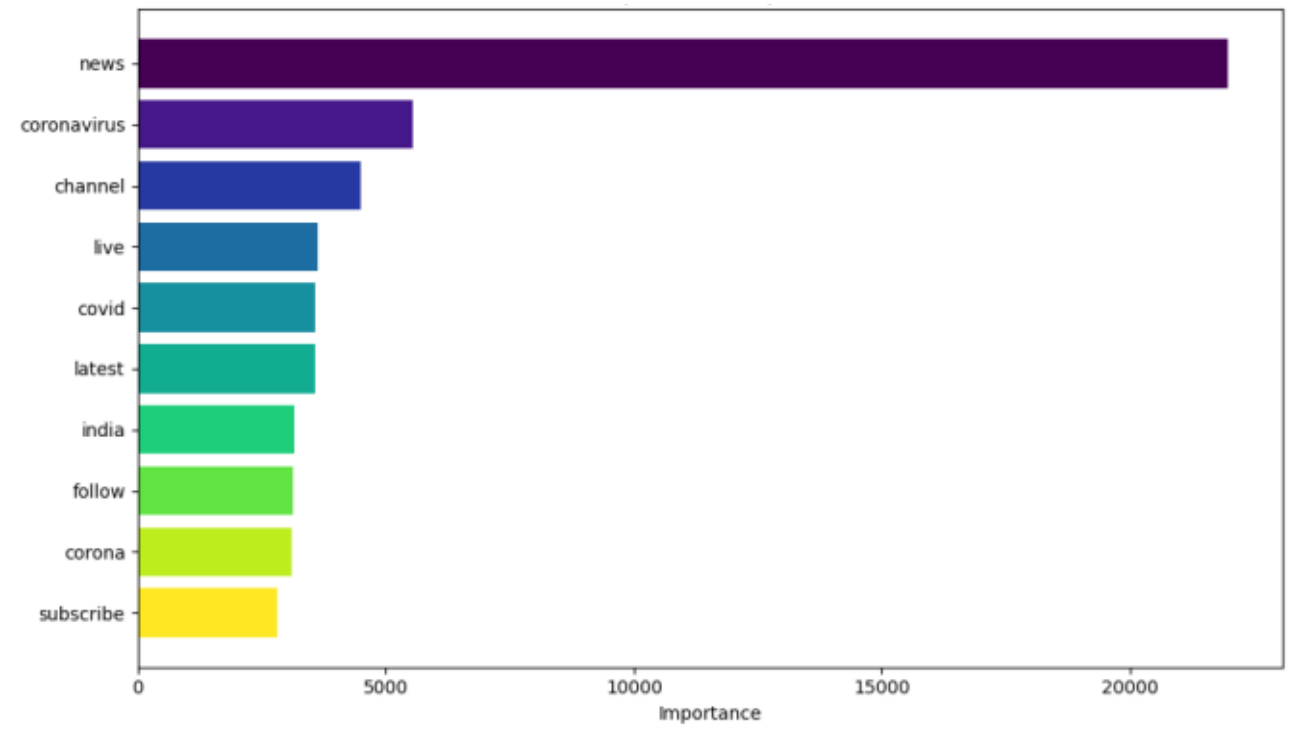

Figure 7. The top 8 frequent words in Topic 1 that focused on real-time news coverage and updates

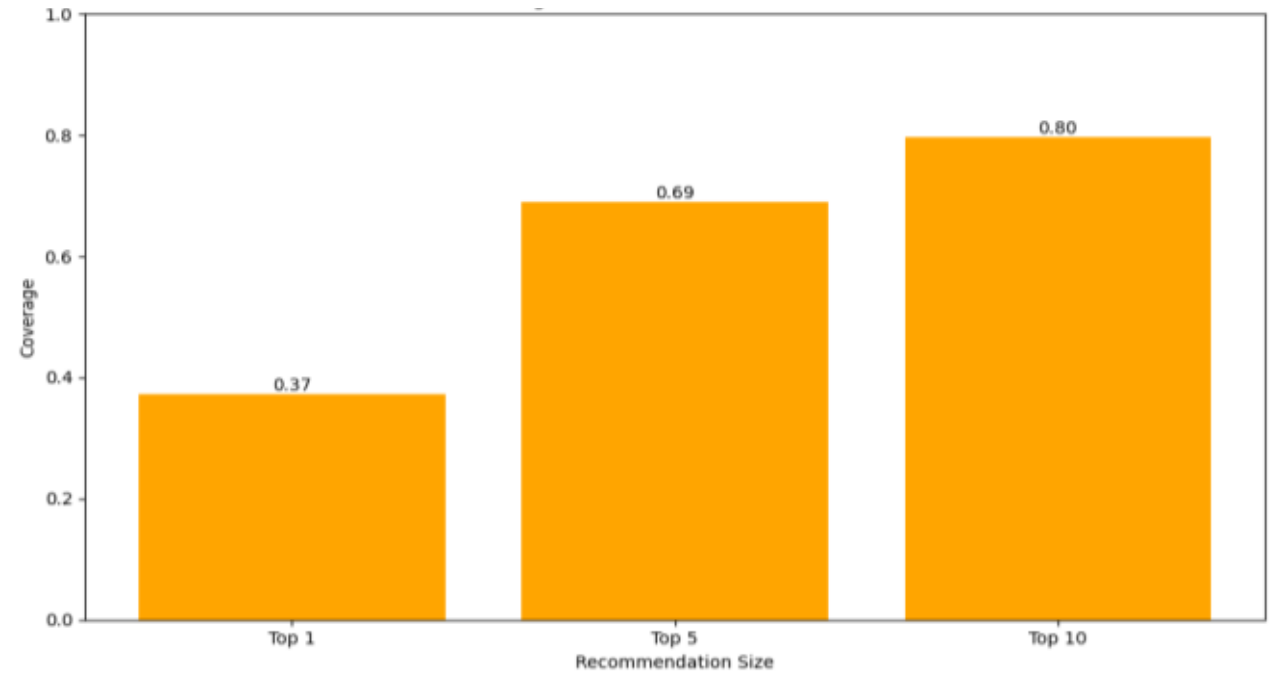

Figure 8. Result of evaluating coverage for different recommendation sizes (top 1, 5, and 10)

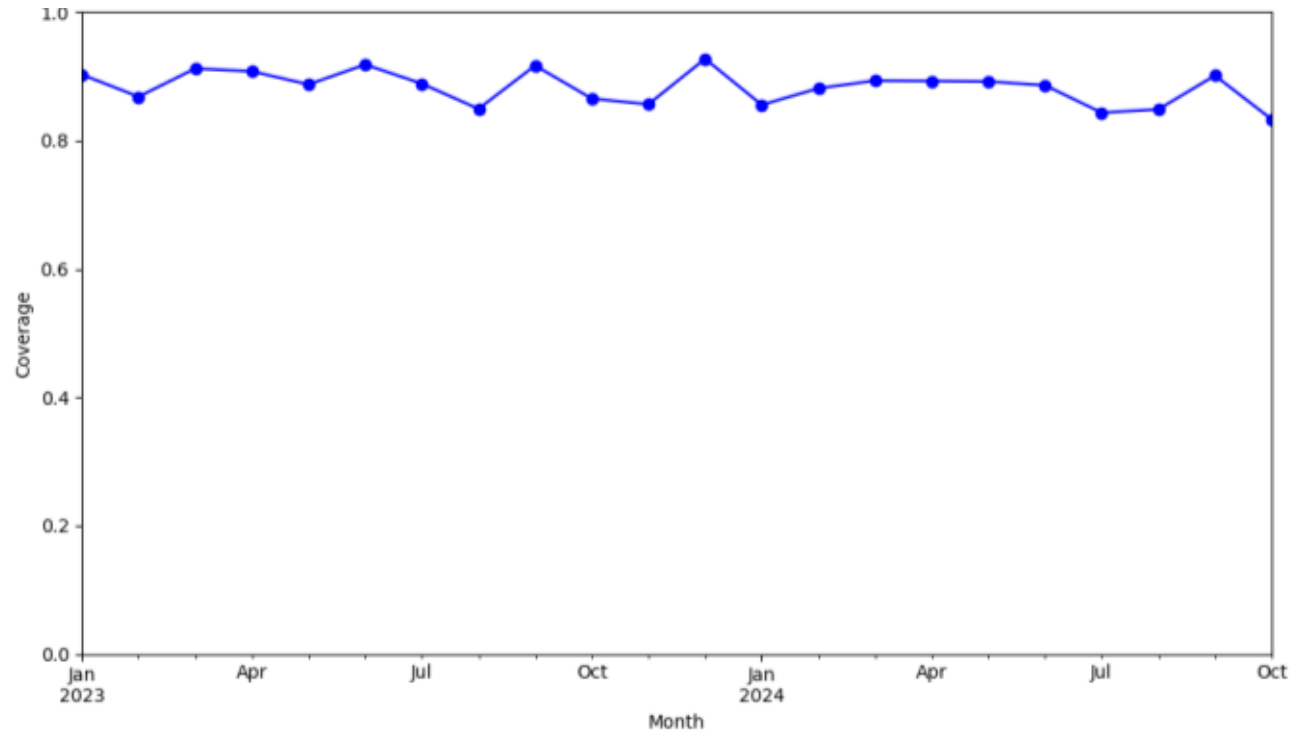

Figure 9. Result of analyzing coverage of recommendations over time (January 2023 to October 2024)

In conclusion, the results of Figures 8 and 9 indicate the effectiveness of a sentiment, toxicity, and topic-aware recommendation framework in navigating COVID-19-related content on YouTube. The recommendation system's high coverage metrics demonstrate its capacity to deliver relevant content related to COVID-19, further highlighting YouTube's role in disseminating pandemic-related information. The sentiment and toxicity analyses (Figures 2 and 3) showed a positive and largely non-toxic environment, underscoring YouTube's potential as a constructive platform for public health discourse. The distinct topics identified through LDA (Figures 5-7) revealed that content creators addressed general and timely informational needs, meeting diverse audience expectations. The work presented in this paper has a couple of limitations. First, the results presented in this paper are based on the data that was collected for this research project. As COVID-19 continues to impact public health on a global scale, influencing and impacting content creation on YouTube [81,82] as well as the views and reactions of the public toward this content [83,84], it is possible that if new data is collected in the near future and this study is repeated, the results obtained might vary as compared to the results presented in this paper. Second, the results from VADER and Detoxify were not manually verified. VADER and Detoxify, while advanced, may not fully capture the complexity of human emotions and toxicity, especially in the context of evolving language use related to COVID-19 on social media platforms such as YouTube, which could lead to incorrect classification on a few occasions.

## V. Conclusion

This paper presents a comprehensive dataset of 9,325 YouTube videos related to COVID-19, spanning from January 1, 2023, to October 25, 2024. The dataset contains rich metadata such as video URLs, video IDs, titles, descriptions, view counts, and other engagement metrics (likes, comments, and duration). The analyses conducted on this dataset - sentiment and toxicity assessment, topic modeling, and recommendation system evaluation, revealed multiple novel insights into how COVID-19-related content on YouTube is structured, presented, and received by viewers. The sentiment analysis showed a predominantly positive or neutral tone across video descriptions, with almost half of the videos categorized as positive. This sentiment distribution suggests that many content creators probably communicated reassuringly, potentially addressing public needs for clarity and positivity amidst a prolonged pandemic. The toxicity analysis underscored YouTube's role as a relatively safe platform for COVID-19 discourse, with over 99% of videos classified as non-toxic. Although a small subset contained potentially toxic content, the low toxicity levels affirm YouTube's suitability for pandemic-related content that aims to educate and inform the general public. Topic modeling using Latent Dirichlet Allocation identified two primary themes: general COVID-19 and health discussions and real-time news and updates. These themes aligned with common viewer needs for general health information and timely updates on the pandemic's progression. The sentiment, toxicity, and topic-based recommendation system, based on TF-IDF vectorization and cosine similarity, proved effective, with 69% coverage across the dataset. The results demonstrate both temporal consistency and scalability, indicating its utility for larger datasets and dynamic content recommendation needs on YouTube. This recommendation system's successful development and testing underscore the value of sentiment, toxicity, and topic-aware video recommendations in enhancing viewer experiences and providing safe and varied content options tailored to different phases and facets of the COVID-19 pandemic. Future work would explore advanced NLP techniques and adaptive recommendation algorithms to further refine recommendation accuracy, particularly for large-scale YouTube datasets on different subject matters.